\documentclass[journal=apchd5,manuscript=article]{achemso}

\usepackage[version=3]{mhchem} 
\usepackage[T1]{fontenc}       

\author{G. Pellegrini}
\email{giovanni.pellegrini@polimi.it}
\author{M. Celebrano}
\author{M. Finazzi}
\author{P. Biagioni}
\affiliation{Dipartimento di Fisica, Politecnico di Milano, Piazza Leonardo da Vinci 32, I-20133 Milano, Italy}

\title[]{Local field enhancement: comparing self-similar and dimer nanoantennas}


\begin{document}

%
%
%
%
%

\begin{abstract}
  We study the local field enhancement properties of self-similar nanolenses and compare the obtained results with the performance of standard dimer nanoantennas. We report that, despite the additional structural complexity, self-similar nanolenses are unable to provide significant improvements over the field enhancement performance of standard plasmonic dimers.
\end{abstract}

Plasmonic nanoantennas are a key ingredient to manipulate light at the nanoscale. Over the last decade, this class of nanostructures has received increasing attention because of their peculiar light-tailoring properties. Different effects including local field and emission enhancement, polarization rotation, angular emission redistribution, nonlinear emission enhancement and high-efficiency single-photon collection are obtained by employing a variety of configurations ranging from plasmonic crystals to tightly coupled nanostructures such as bow-tie and Yagi-Uda antennas\cite{Novotny_2011,Biagioni_2012}. Despite the considerable variety of attainable effects, the generation of large local field enhancements remains a crucial factor for the successful design of a large class of devices such as sensing, light emission, and non-linear plasmonic platforms. This has led to the realization of a vast range of nanostructures, mostly exploiting the presence of a geometrical gap, to achieve field enhancements of about 2 orders of magnitude in amplitude\cite{Novotny_2011,Biagioni_2012}. Self-similar plasmonic nanolenses represent a notable exception in the plasmonic nanoantenna landscape, employing cascade field enhancement amplification to nominally achieve, in ideal conditions, unprecedented field amplitude gains of more than 3 orders of magnitude\cite{Li_2003}. This unique field enhancement ability has led to a significant interest in self-similar nanostructures, and likewise to a close scrutiny of their optical properties\cite{Li_2006,li_li_2006,Dai_2008}, finally giving impulse to several research efforts aiming to fabricate and employ this class of geometries\cite{Li_2005,Li_2005a,Bidault_2008,Ding_2010,Kravets_2010,Toroghi_2012,Toroghi_2012a,H_ppener_2012,Almpanis_2012,Gramotnev_2013,Coluccio_2015}.
In this paper we investigate the field enhancement properties of self-similar plasmonic nanolenses, and compare them with those of standard plasmonic gap antennas such as sphere, rod and bowtie dimers, featuring similar geometrical parameters. We aim to unravel the physical mechanisms contributing to the extreme field enhancement properties reported for self-similar nanolenses, with particular attention to the effect of radiative and non-radiative losses on the obtained local field enhancement.
\begin{figure}[t!]
\begin{center}
\includegraphics[]{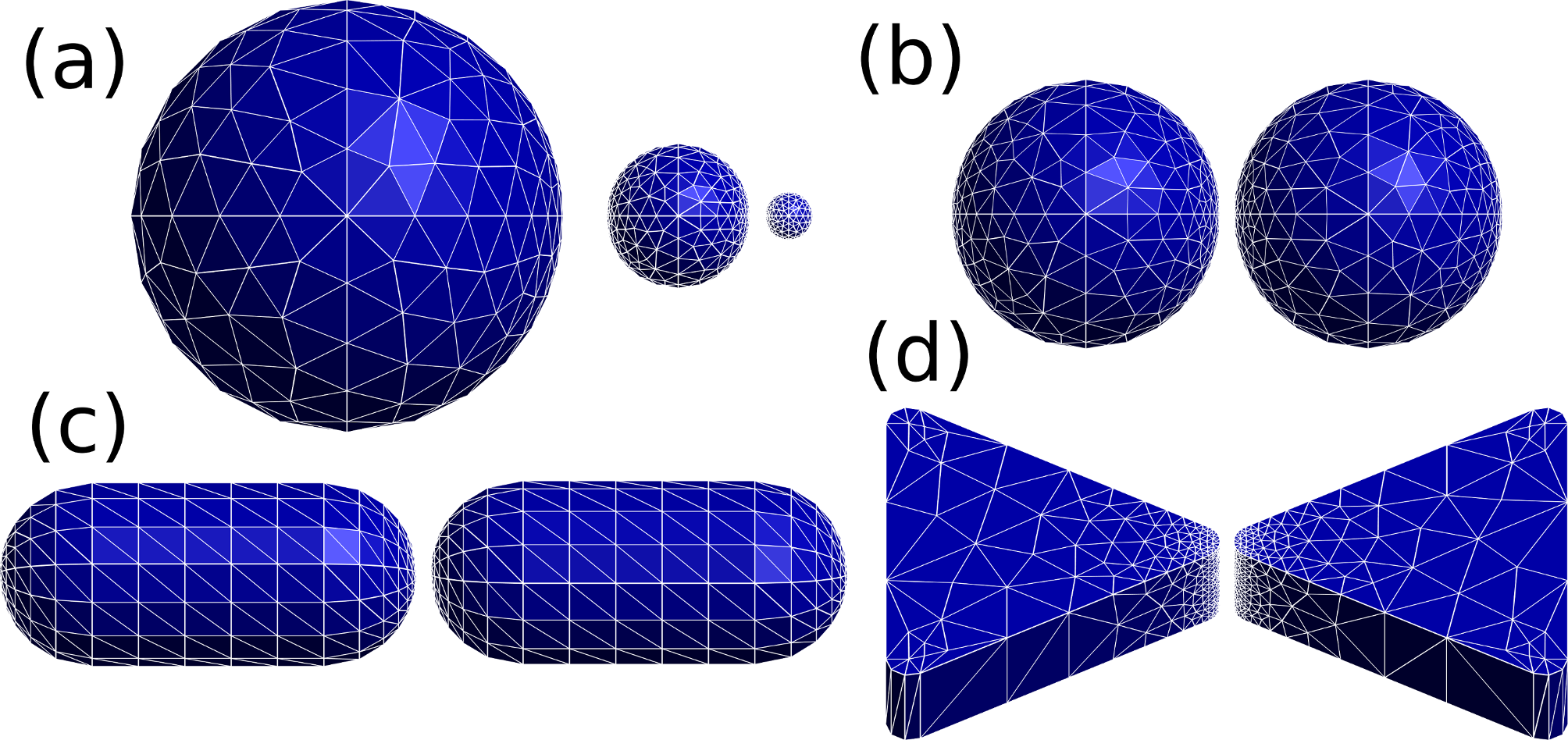}
\caption{\label{Fig1} Surface meshes, adopted for the boundary element method calculations, for each of the modeled nanostructures. (a) Self-similar nanolens; (b) Sphere dimer; (c) Rod dimer; (d) Bowtie antenna.%
}
\end{center}
\end{figure}

We employ a Boundary Element Method (BEM) approach\cite{Reid_2015,Reid_scuffem_2016}, complemented by Generalized Multiparticle Mie (GMM) calculations for spherical geometries\cite{Pellegrini_2007,Pellegrini_pygmm_2016}, to study the local field enhancement properties of the self-similar and dimer plasmonic structures sketched in Fig.\ref{Fig1}. We illuminate the antennas with a monochromatic plane wave in which the electric field vector is parallel to the antenna principal axis and the wavevector is normal to the structure horizontal symmetry plane. The self-similar antenna is defined by the radius $R_{i}$ of each sphere and the respective sphere to sphere separation $d_{i,i+1}$. Defining $\kappa$ as the self-similarity factor, i.e. the scaling factor between two successive elements of the array, we have $R_{i+1}=\kappa R_{i}$ and $d_{i+1,i+2}= \kappa d_{i,i+1}$. In the present case we take $\kappa=\frac{1}{3}$ and define the interparticle spacing as $d_{i,i+1}=0.6 R_{i+1}$.
Choosing the largest feature size as $R_{1}=45$~nm we obtain $R_{2}=15$~nm and $R_{3}=5$~nm, with the smallest separation equal to $d_{2,3}=3$~nm, i.e a self-similar nanolens corresponding to the first investigated structure in Ref.\citenum{Li_2003}. The comparison of the field enhancement properties among the different nanostructures dictates the choice of the same gap size $g=d_{2,3}=3$~nm for all the standard plasmonic dimers. Furthermore, in order to perform a fair comparison, we introduce two additional geometrical constraints for the design of the dimer antennas, one on the antenna volume, with $V_{dimer} \leq V_{nanolens}$, and the other on the longitudinal dimension (i.e. along the polarization direction of the illumination field) of each antenna element, with $L_{element} \leq 2 R_{1}$.
We finally adopt a sphere dimer with radius $R_{sphere}=30$~nm (Fig.\ref{Fig1}(b)), a cylindrical rod dimer capped with two hemispheres with total rod length $L_{rod}=90$~nm and radius $R_{rod} = 20$~nm (Fig.\ref{Fig1}(c)), and a bowtie antenna with side $L_{bowtie}=90$~nm, height $h_{bowtie}= \frac{ \sqrt{3}}{2} L_{bowtie}$ (equilateral triangle), thickness $t_{bowtie}=20$~nm and corner curvature radius $r_{bowtie}=5$~nm (Fig.\ref{Fig1}(d)),corresponding to the self-similar nanolens smallest geometrical feature. It is finally worth noting that the adopted geometrical parameters allow us to exclude, in a first approximation, tunneling and non-locality effects at the gap\cite{Zuloaga_2009,Esteban_2012,Savage_2012}, and to use bulk silver optical constants retrieved from the literature\cite{Johnson_1972,Yang_2015,Palik1985}. Furthermore, we ignore the size corrections for the electron mean free path for the $R_{3}=5$~nm nanolens sphere\cite{Li_2003}.

\begin{figure}[t!]
\begin{center}
\includegraphics[]{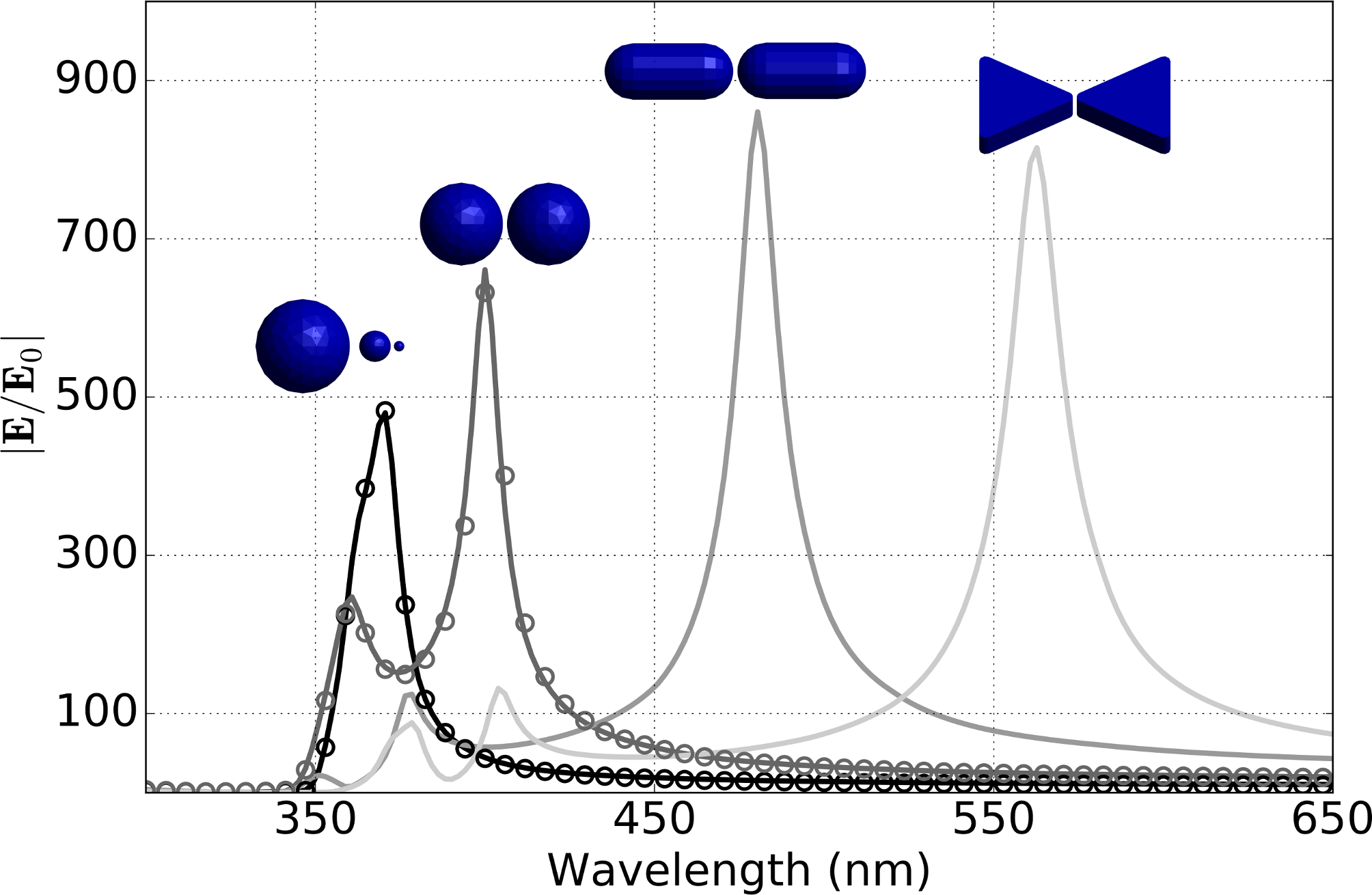}
\caption{\label{Fig2} Field enhancement calculated with BEM method in quasi-static approximation. Fields are monitored in the middle of the gap in the case of standard dimers, and $0.5$~nm from the smallest sphere surface in the case of the self-similar nanostructure. Dots: field enhancement spectra calculated with GMM approach.%
}
\end{center}
\end{figure}

Figure~\ref{Fig2} reports local field enhancement spectra for the silver self-similar nanolens and the standard dimer antennas, where the optical constant of silver is retrieved from the literature\cite{Johnson_1972,Li_2003}. Fields are computed in quasi-static approximation and are monitored in the middle of the gap in the case of the standard dimers, and on the simmetry axis $0.5$~nm from the smallest sphere surface in the case of the nanolens, to take into account the highly asymmetrical field distribution at the gap\cite{Li_2003}. The self-similar nanolens and the sphere dimer display field enhancements in the same order of magnitude, where the sphere dimer shows a slightly more intense and red-shifted resonance at about $400$~nm. The field enhancement spectra, calculated both by BEM and GMM approaches, are in excellent quantitative agreement, therefore indicating that the choice of mesh resolution for BEM calculations, and the number of included multipole expansions in the case of the GMM approach ($n_{\mathrm{stop}}=20$), are appropriate\cite{li_li_2006}. Field spectra for rod dimer and bowtie antenna exhibit similar trends, with larger peaks intensity and red-shifted resonance position with respect to the nanolens.
Overall, the computed field spectra clearly show that standard antenna dimers can produce field enhancements comparable or larger to those obtained employing self-similar nanostructures, even though the gap geometry, as a general rule, imposes a red shift of the dipolar resonance with respect to that of a self-similar antenna, which in first approximation appears at the single sphere energy.

\begin{figure}[h!]
\begin{center}
\includegraphics[]{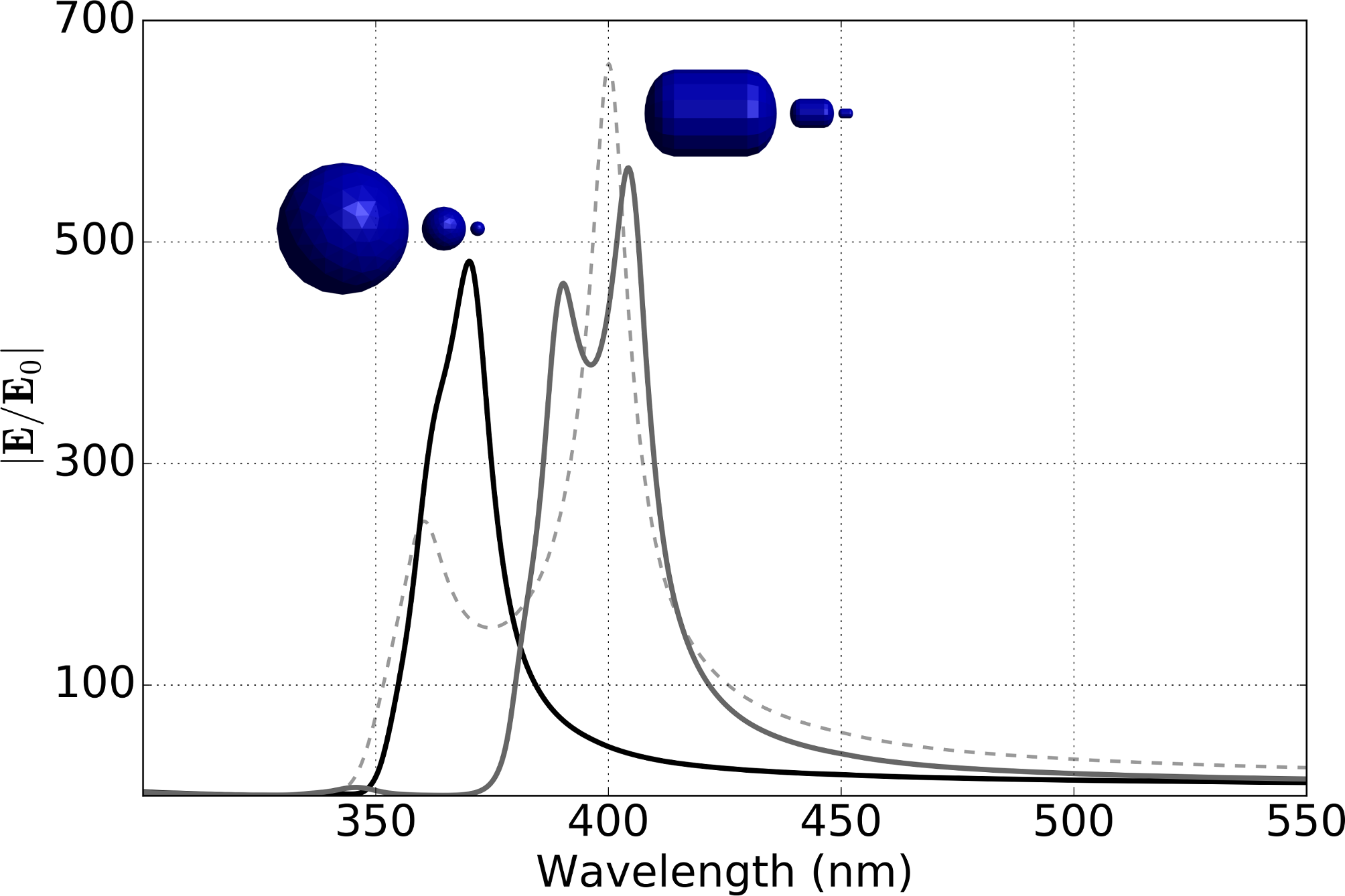}
\caption{\label{Fig3} Field enhancement spectra calculated in quasi-static approximation for sphere and rod self-similar nanolenses. Lateral dimensions, gaps, and surface curvature radii of the rod nanolens are equal to those of the sphere structure. Fields are monitored $0.5$~nm from the smallest particle surface. Dashed line: field enhancement spectrum for sphere dimer antenna.%
}
\end{center}
\end{figure}

It is now necessary to investigate the role of the resonance wavelength on the field enhancement properties of the self-similar array. To do so we design a self-similar rod nanostructure resonating at $400$~nm, which also corresponds to the resonance of the sphere dimer antenna. In order to properly target the effects of the wavelength dependence, lateral dimensions, gaps, and self-similarity constant $\kappa$ in the rod array are chosen to be identical to those found in the self-similar spherical geometry, with the first rod radius set as $R_{1,rod}=30$~nm. Furthermore, in this case, we adopt an oblate ellipsoidal capping for the rods, with major and minor axis length defined as $a_{1,rod}=30$~nm  and $b_{1,rod}=20$~nm respectively, resulting in a curvature radius of $r_{1,c}=\frac{a^{2}}{b}=45$~nm, i.e. identical to that found in the spherical geometry.
Local field spectra reported in Fig.\ref{Fig3} indeed verify that, even if a minor effect due to the wavelength dependence can be observed as an adjustment from $|\mathbf{E}/\mathbf{E}_{0}| \sim 500$ to $|\mathbf{E}/\mathbf{E}_{0}| \sim 550$, self-similar nanolenses are unable to provide larger field enhancements than those obtained employing standard plasmonic dimers within the adopted geometrical constraints and, in this specific case, the self-similar rod geometry provides smaller enhancements than a simple sphere dimer with an equal gap.
\begin{figure}[t!]
\begin{center}
\includegraphics[]{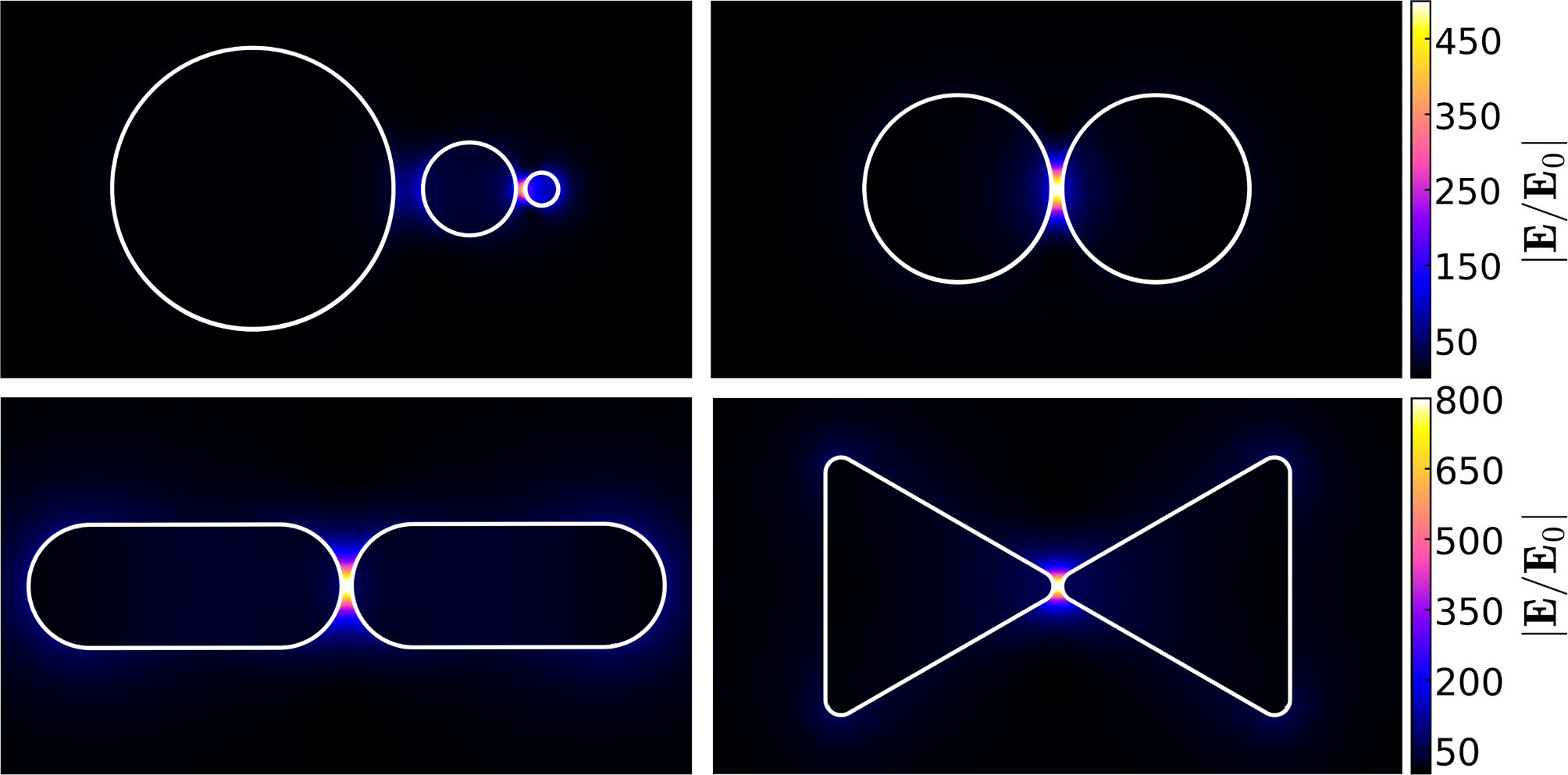}
\caption{\label{Fig4} Field amplitude enhancement mappings. Mappings are computed at the resonance wavelength and at the equatorial plane of each nanostructure. (a) Self-similar nanolens; (b) Sphere dimer; (c) Rod dimer; (d) Bowtie antenna.%
}
\end{center}
\end{figure}
Local field mappings computed at resonance, and reported in Fig.\ref{Fig4}, return a picture coherent with that of the spectral calculations, with local intensity profiles in the gap of the self-similar structure in very good agreement with those reported in Ref.\citenum{Li_2003}. In all cases standard plasmonic antennas provide stronger and more uniform local-field hot-spots, whereas the only advantages of self-similar nanostructures lie in a slightly better spatial confinement, and in the ability to sustain resonances in the blue end of the spectrum.
This first analysis returns a clear picture where realistic nanolens geometries, in ideal quasi-static conditions, are unable to provide field enhancements much larger that those obtained with standard plasmonic dimers. A natural conclusion is that, in the considered scenario, cascade amplification does not justify the observed field enhancement values for self-similar nanostructures, and that the mechanism responsible for the observed extreme field enhancement must be operating in each of the analyzed nanostructures. A likely explanation for the observed behavior can probably be found in the adoption of the quasi-static approximation, which completely eliminates the presence of radiative losses, and likewise in the choice of silver as adopted material, which significantly limits ohmic dissipation\cite{Li_2003,Johnson_1972}.

We first investigate the role of ohmic dissipation on the field enhancement properties of self-similar and standard plasmonic dimers. To this end we calculate field enhancement spectra adopting three different optical constants for silver, each one characterized by distinct imaginary parts of the dielectric function, and as a consequence by different ohmic losses\cite{Johnson_1972,Palik1985,Yang_2015}.

\begin{figure}[t!]
\begin{center}
\includegraphics[]{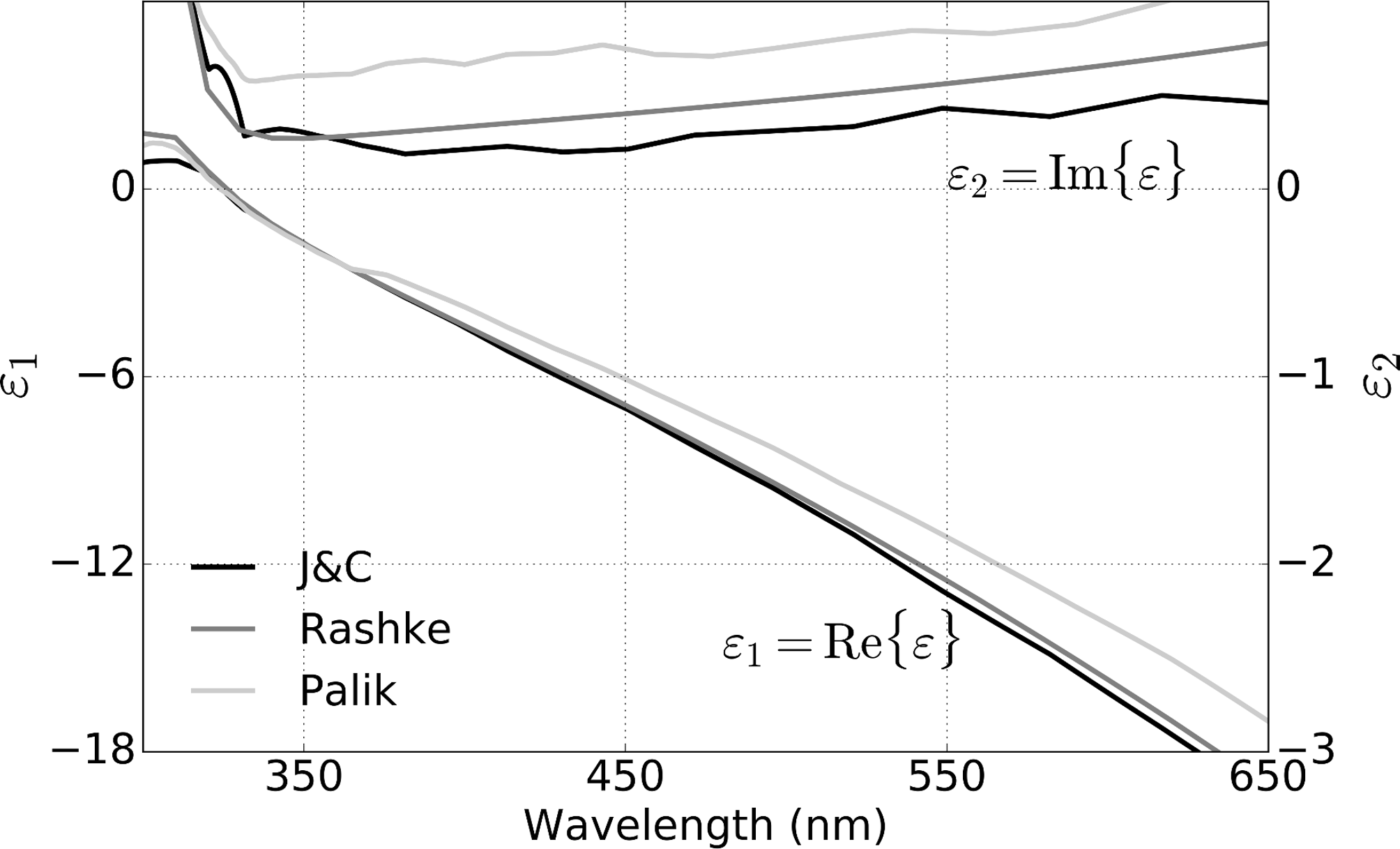}
\caption{\label{Fig5} Dielectric constant real ($\varepsilon_{1}$) and imaginary part ($\varepsilon_{2}$) for each of the adopted silver optical constants.}
\end{center}
\end{figure}

\begin{figure}[t!]
\begin{center}
\includegraphics[]{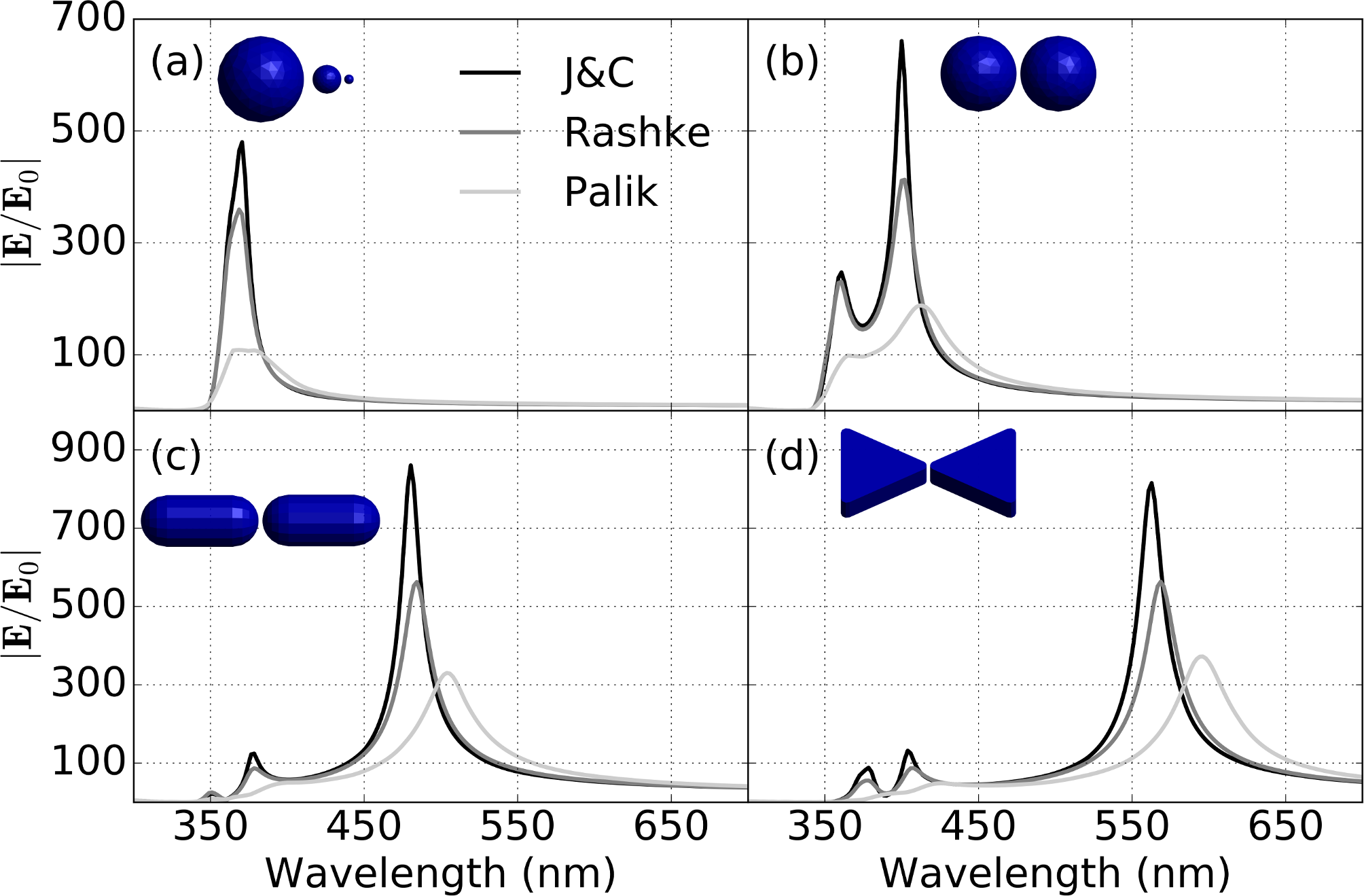}
\caption{\label{Fig6}  Field enhancement spectra calculated in quasi-static approximation. Enhancement spectra are calculated for all adopted optical constants for each nanostructure. (a) Self-similar nanolens; (b) Sphere dimer; (c) Rod dimer; (d) Bowtie antenna.%
}
\end{center}
\end{figure}

Figure~\ref{Fig5} reveals how the optical constant retrieved from Ref.\citenum{Johnson_1972}, which is also the one adopted in Ref.\citenum{Li_2003} and in the performed quasi-static calculations, is indeed characterized by the smallest imaginary part $\varepsilon_{2}$ of the dielectric function, and therefore likely to provide the largest field enhancements. This hypothesis is confirmed by the field enhancement spectra displayed in Fig.\ref{Fig6}; the adoption of optical constants characterized by distinct ohmic losses results in major differences in the calculated local field enhancements. In the case of self-similar nanostructures, values obtained adopting the Johnson and Christy optical constants are up to 5 times larger than those  computed using the alternative dielectric functions, while the peak difference diminishes to a factor of 3 in the case of standard dimers. The difference in variability can be partially explained by the different spectral properties of the adopted materials, nevertheless a further contribution derives from the cascade amplification mechanism which, being expressed in first approximation as $g_{n} \sim {(\frac{\varepsilon_{1}}{\varepsilon_{2}})}^{n}$, with $n$ the number of cascade steps,  shows a strong dependence on the imaginary part of the optical constant\cite{Li_2003}.

\begin{figure}[t!]
\begin{center}
\includegraphics[]{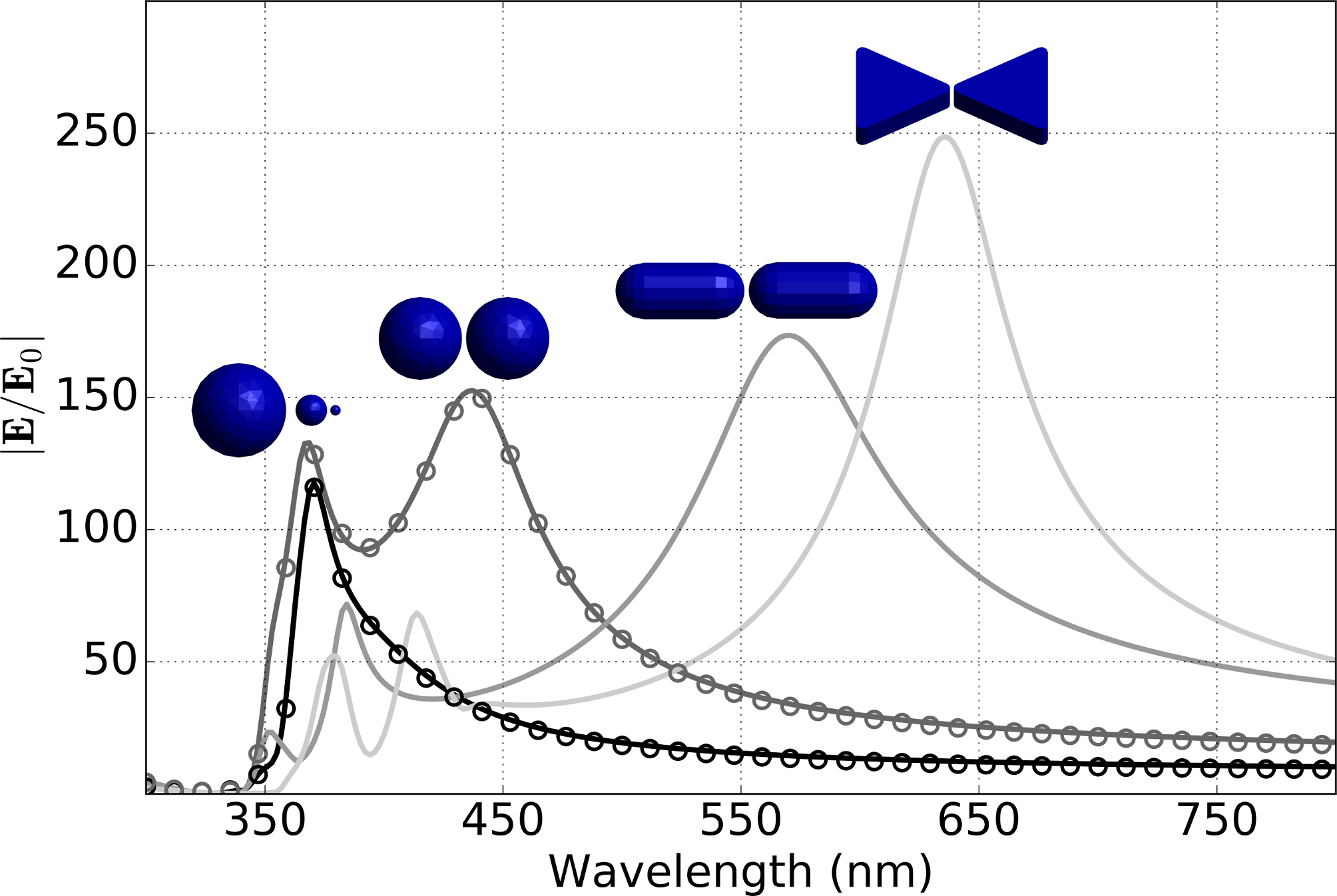}
\caption{\label{Fig7} Field enhancement spectra calculated including full retardation. Fields are monitored in the middle of the gap in the case of standard dimers, and $0.5$~nm from the smallest sphere surface in the case of the self-similar nanostructure. Dots: field enhancement spectra calculated with GMM approach.%
}
\end{center}
\end{figure}

Finally, in order to investigate the role of radiative losses and provide a realistic estimate of the field enhancement achievable with self-similar and standard plasmonic dimers, we perform a local field spectral analysis including full retardation effects, this time adopting a silver optical constants featuring intermediate ohmic losses\cite{Yang_2015}. The introduction of full retardation in this size regime causes a further reduction of peak local field enhancements, as well as significant red-shift and broadening of all the dipolar resonances, as displayed in Fig.\ref{Fig7}. Both the GMM and BEM calculated spectra for the self-similar nanolens and the sphere dimer are in excellent quantitative agreement, with peak amplitudes reaching $|\mathbf{E}/\mathbf{E}_{0}| \sim 150$ for the dimer, and $|\mathbf{E}/\mathbf{E}_{0}| \sim 100$ for the self-similar nanostructure. The introduction of radiative losses thus induces an additional reduction of a factor of 3 in field peak values, which is also reflected in the trends observed for rod dimers and bowtie antennas. In this case a slight exception is represented by the bowtie structure, whose maximum fields are less affected by the opening of radiative channels than the standard and self-similar counterparts. Overall, the introduction of radiative and ohmic losses can lead to a reduction in field enhancement up to one order of magnitude in a worst case scenario.

In conclusion, we have studied and compared the field enhancement properties of self-similar and standard plasmonic dimer nanostructures. In this context we have shown that self-similar nanolenses are unable to provide, even in ideal quasi-static conditions, local field enhancements significantly larger than those obtained employing standard plasmonic dimers. We have also highlighted the influence of radiative and ohmic losses on the final field enhancement performance, and established that the observed extreme enhancements likely descend from the absence of radiative losses, intrinsic to the quasi-static approximation, and from the adoption of materials featuring extremely small ohmic losses. Furthermore, we have also shown that an appropriate inclusion of all loss channels results in a field enhancement reduction up to one order of magnitude in a worst case scenario. Finally, we have demonstrated that the adoption of self-similar nanostructures to achieve large local field enhancements, with the exception of the blue range of the visible spectrum, does not seem justified, and that standard plasmonic dimers offer the best tradeoff in terms of performance and fabrication feasibility.

\begin{acknowledgement}

The research leading to these results has received funding from
the European Union's Seventh Framework Programme under
grant agreement no. 613055.
The authors declare no competing financial interest.

\end{acknowledgement}


\bibliography{bibliography}

\end{document}